\documentclass[twocolumn,showpacs,preprintnumbers,amsmath,amssymb,footinbib]{revtex4}

\usepackage{graphicx}
\usepackage{dcolumn}
\usepackage{bm}

\begin{document}

\preprint{APS/123-QED}
\title{Magneto-thermal evidence of a partial gap at the Fermi surface of UPt$_2$Si$_2$}

\author{N. Johannsen$^{1*}$, S. S\"{u}llow$^{2}$, A. V. Sologubenko$^1$, T. Lorenz$^1$,  and J. A. Mydosh$^1$}

 \affiliation{$^1$II. Physikalisches Institut, Universit\"{a}t zu K\"{o}ln, Z\"{u}lpicher Str. 77, 50937 K\"{o}ln, Germany\\
$^2$Institut f\"{u}r Physik der kondensierten Materie, TU Braunschweig, 38106 Braunschweig, Germany}

\date{\today}

\begin{abstract}
Motivated by the observation of a giant Nernst effect in URu$_2$Si$_2$, the thermoelectric response of the related system UPt$_2$Si$_2$ was investigated using thermal and electric transport properties such as the Nernst and Seebeck effects, thermal conductivity, Hall effect and electrical resitivity. Unlike URu$_2$Si$_2$, UPt$_2$Si$_2$ is neither superconducting nor exhibits a ``hidden-order'' state. Nevertheless a pronounced Nernst effect anomaly is found to coincide with the onset of the antiferromagnetic order in UPt$_2$Si$_2$. Although the absolute values are substantially lower, its appearance and characteristics can favorably be compared to the giant Nernst effect in URu$_2$Si$_2$ indicating the common feature of a partial Fermi surface gap.
\end{abstract}

\pacs{71.27.+a, 72.15.Jf, 87.15.Zg}

\maketitle

The mysterious nature of the ``hidden-order'' (HO) phase in URu$_2$Si$_2$, discovered more than 20 years ago, is still unresolved \cite{palstra85,schlabitz86,maple86}. Initially HO was associated with a weak antiferromagnetic order setting in at $T_{HO}\simeq$ 17.5\,K \cite{broholm91,broholm87}, however recent experiments have shown this transition to be essentially non-magnetic \cite{amitsuka07}. Transport properties such as the Seebeck effect or the Nernst and Hall effects are very sensitive in detecting slight changes in the charge carrier spectrum at the Fermi energy. The appearance of a giant Nernst effect \cite{bel04} emerging with the onset of the HO impelled the investigation of these transport properties in the related system UPt$_2$Si$_2$ which forms a local-moment antiferromagnetic ground state \cite{suellow08}. 

While for the high-temperature superconductors \cite{xu00,wang06,johannsen07} the anomalous part of the  Nernst effect is caused by moving flux lines, other mechanisms have to be held responsible for  systems comprising the various classes of heavy fermions \cite{bel04,bel04a,onose07}, semimetals \cite{behnia07a}, ferromagnetic metals \cite{miyasato07,lee04,taguchi01} or organic compounds \cite{nam07,wu03}.
Among these are anomalous scattering mechanisms like ``side jump'' \cite{berger70} or ``skew scattering'' \cite{smit55}. They are driven by magnetic order and are sources to the anomalous components of the Hall effect. Furthermore in some systems Nernst effects can arise due to a combination of unusual physical properties. According to Oganesyan and Usshishkin \cite{oganesyan04}, a large Nernst signal can arise in systems that combine lightweight charge carriers, long scattering lifetimes and reduced Fermi energies. Not only the latter but also a gap or a partial gapping of the Fermi surface (FS) has been identified as the source of anomalies in the Nernst effect \cite{dik75} and the thermoelectric power \cite{abelskii72} in antiferromagnets. These calculations consider the presence of a superzone gap in addition to phonon and spin scattering mechanisms. 

In this Letter, we present thermal and electric transport properties such as the Nernst and Seebeck effects, thermal conductivity, Hall effect and electrical resitivity. The predictions that arise from the calculations of \cite{abelskii72,dik75} include substantial changes in the Nernst, Seebeck and Hall effects with their interrelated scaling and a maximum formation at about $0.5\cdot T_N$. All these features are in very good agreement with the measurements presented here, indicating that gap formation strongly influence these transport properties. However, the considered gapping of the FS can affect only parts of it because the resistivity shows no sign of a metal to insulator transition. Since such a partial gapping has also been observed for URu$_2$Si$_2$ at the onset of the HO \cite{jeffries07,jo07}, the appearance of the giant Nernst effect might as well be discussed as being caused by a Fermi surface reconstruction. Our measurements clearly demonstrate that the appearance of a Nernst signal tracks the phase transition at $T_N\simeq$ 32\,K and its behavior is qualitatively comparable to the effects reported for the HO transition of URu$_2$Si$_2$.

 The structure of UPt$_2$Si$_2$ is tetragonal with lattice constants $a=$\,4.186\,{\AA} and $c=$\,9.630\,{\AA}. The lattice is of CaBe$_2$Ge$_2$ type (space group $P$4/$nmm$). UPt$_2$Si$_2$ possesses an antiferromagnetic ground state with $T_N\simeq$ 32\,K \cite{steeman90}. Neutron diffraction studies revealed magnetic moments of 2.5\,$\mu_B$ oriented along the $c$ direction with ferromagnetic coupling within the $a-a$ planes and antiferromagnetic coupling on adjacent planes along the $c$ axis. 

The thermal and electric transport measurements have been carried out on two annealed samples of UPt$_2$Si$_2$. The as-cast samples were annealed for approximately one week at a temperature of $T=$900$^\circ$C. One sample was oriented such that the application of either a temperature gradient or a transport current was along the $a$ axis with the magnetic field applied along the $c$ axis and the other sample with the transport quantities applied along $c$ which leaves the magnetic field and the transverse components in the $a-a$ plane. Temperature gradients have been applied using a chip heater glued to the top of the sample and were picked up by AuFe-Chromel thermocouples. The setup in which a considerable Nernst signal, $e_N=E_y/(-\nabla_xT)$, could be detected was oriented such that $\nabla T \parallel a$ and $H\parallel c$. In $\nabla T \parallel c$ and $H\parallel a$, no Nernst signal could be resolved so that we will continue focussing upon the former configuration.
\begin{figure}
\center
\includegraphics[width=7.5cm, clip]{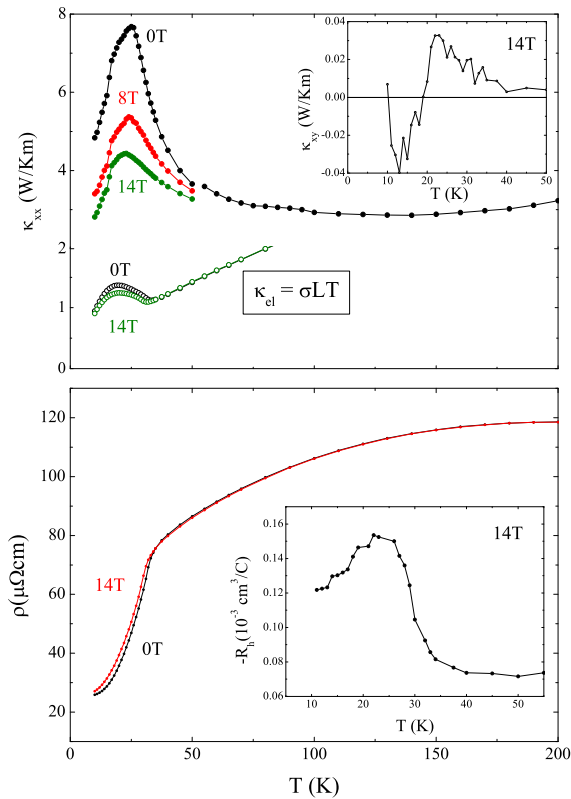}
\caption{(Color online) Upper panel: thermal conductivity along the $a$ axis of UPt$_2$Si$_2$ for magnetic fields applied along $c$. The inset shows the off-diagonal component $\kappa_{xy}$, which is of the order of 1\% of $\kappa_{xx}$. The open symbols give an upper-limit estimation of the electronic contribution to the thermal conductivity as calculated by the Wiedemann-Franz law. Lower panel: Resistivity along the $a$ axis in 0 and 14\,T along $c$. The inset shows the temperature dependence of the Hall coefficient of UPt$_2$Si$_2$ in 14\,T with its pronouced increase below $T_N$.}\label{fig:kappa_temp_wied_el}
\end{figure}
Figure~\ref{fig:kappa_temp_wied_el} (lower panel) shows the temperature dependence of the resistivity of UPt$_2$Si$_2$ along the $a$ axis with the magnetic field applied along the $c$ axis. Coming from high temperatures $\rho (T)$ shows a very weak temperature dependence with a broad maximum of $\rho (T)\approx 120\mu\Omega$cm around 200\,K.  At $T_N \simeq 31.5$\,K (determined from the maximum of $d\rho/dT$) the transition into the antiferromagnetic phase shows up as a kink in $\rho$ with no signature of a superzone gap. The knee-like reduction of $\rho(T)$ can be modeled within a scenario in which magnetic excitations freeze out due to the opening of a spin-wave gap with $\Delta \approx 44$\,K \cite{andersen80,suellow08}. Electronic transport shows localized behavior along the $c$ axis which near $T_N$ exhibits a clear dip in $d\rho/dT$ denoting the opening of a superzone gap in the charge channel (see Fig.~8 of Ref\,\cite{suellow08}).

The upper panel of Fig.~\ref{fig:kappa_temp_wied_el} displays the temperature dependence of the thermal conductivity in zero field, 8\,T, and 14\,T.
For $H=0$, $\kappa(T)$ demonstrates a weak temperature dependence  at high temperatures, with a shallow minimum at $T \approx 125$\,K. With decreasing temperature, $\kappa(T)$ develops a peak at about 25\,K. 
The height of the peak is strongly $H$-dependent, while its position changes with $H$ only very weakly.  
The inset of the upper panel of  Fig.~\ref{fig:kappa_temp_wied_el} displays 
the off-diagonal component of the thermal conductivity, $\kappa_{xy}$, also referred to as Righi-Leduc effect.
The absolute values of $\kappa_{xy}$ are more than two orders of magnitude smaller than the corresponding $\kappa_{xx}$ values. 
The open symbols of Fig.~\ref{fig:kappa_temp_wied_el} present an estimation of the upper limit of the electronic thermal conductivity $\kappa_{\rm el}$ calculated from the experimental data of $\rho(T)$ using the Wiedemann-Franz law, $\kappa_{\rm el}=L T / \rho$ with $L=$\,2.45$\cdot 10^{-8}$\,W/$\Omega$K$^2$ being the Lorenz number. 
At low $T$, the calculated $\kappa_{\rm el}$ is much smaller than the total measured $\kappa$; besides, it does not account for the strong suppression of $\kappa$ by magnetic field. 
This indicates that the thermal conductivity in UPt$_2$Si$_2$ is predominantly phononic.    
The strong sensitivity of thermal conductivity to the magnetic field in a broad temperature region, both above and below $T_N$, suggests that spin fluctuations play an important role as scatterers of phonons. It is surprising, however, that no clear anomaly of $\kappa(T)$ is observed at the magnetic ordering transition, see e.g.\cite{berggold07}.

\begin{figure}
\center
\includegraphics[width=7.5cm, clip]{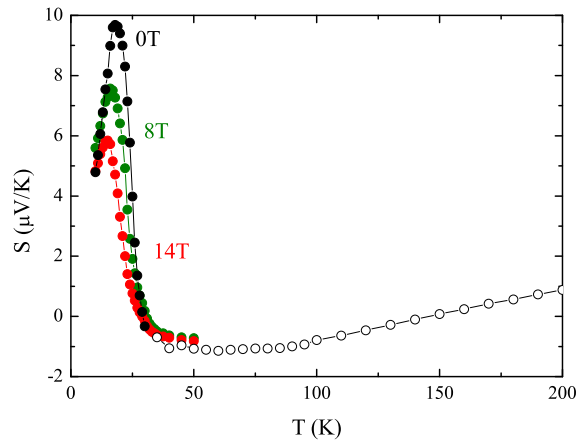}
\caption{(Color online) Thermopower of UPt$_2$Si$_2$ for magnetic fields of 0, 8\,T, and 14\,T ($\nabla T||a$, $H||c$).}\label{fig:thp_temp}
\end{figure}

The onset of the antiferromagnetic ordering is also accompanied with an enhancement of the thermopower (Fig.~\ref{fig:thp_temp}) which furthermore becomes field dependent such that increasing fields suppress $S(T,H)$. The zero-field thermopower exhibits a steep increase in the vicinity of the antiferromagnetic order and peaks at about half $T_N$. Applying a magnetic field leads to a gradual suppression of the absolute values of $S(T\leq T_N)$ and slightly shifts its maximum to lower $T$.

\begin{figure}
\center
\includegraphics[width=7.5cm, clip]{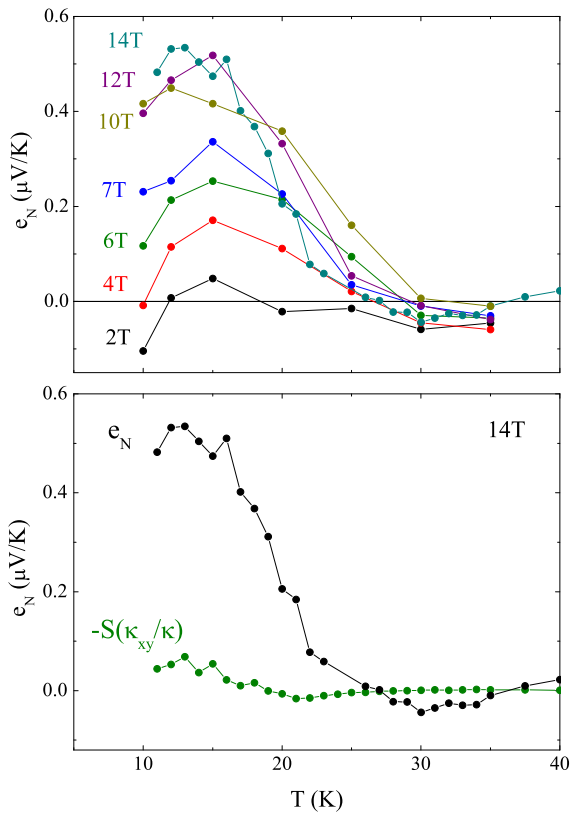}
\caption{(Color online) Upper panel: Temperature dependence of the Nernst signal of UPt$_2$Si$_2$ in various magnetic fields. Lower panel: Comparison of the temperature dependent thermal Hall contribution to the measured Nernst signal, in $H=$\,14\,T ($\nabla T||a$, $H||c$).}\label{fig:nernst_temp}
\end{figure} 
A large Nernst signal emerges below the antiferromagnetic transition temperature of $T_N\simeq$\,32\,K as depicted in Fig.~\ref{fig:nernst_temp}. The signal is not anomalous in a way that it deviates from a field-linear quasiparticle background as is the case in many high-temperature superconductors. The signal here just rises below the temperature of antiferromagnetic order and displays a linear field dependence. Upon lowering the temperature, $e_N$ gradually increases until a maximum at $\approx 15$\,K is reached. Further cooling causes a decrease of the signal.

 In order to fully analyze the various contributions that the adiabatic Nernst signal may be composed of \cite{onose07}, the Seebeck effect, Hall effect and Righi-Leduc effect have been additionally measured as already introduced,
\begin{equation}
e_N=\rho\alpha_{xy}-S\left[\frac{\sigma_{xy}}{\sigma}+\frac{\kappa_{xy}}{\kappa}\right].\label{eq:onose2}
\end{equation}
Here, $\alpha$ and $\sigma=\rho^{-1}$ denote the Peltier and conductivity tensor, respectively. 
 To detect tiny transverse temperature gradients, an AuFe-Chromel thermocouple has been attached directly to the wires that pick up transverse voltages. The difference between the isothermal and the adiabatic Nernst signal is given by the product of the thermopower and $\kappa_{xy}/\kappa$ and is shown in Fig.~\ref{fig:nernst_temp} (lower panel) at 14\,T. 
\begin{figure}
\center
\includegraphics[width=8.3cm, clip]{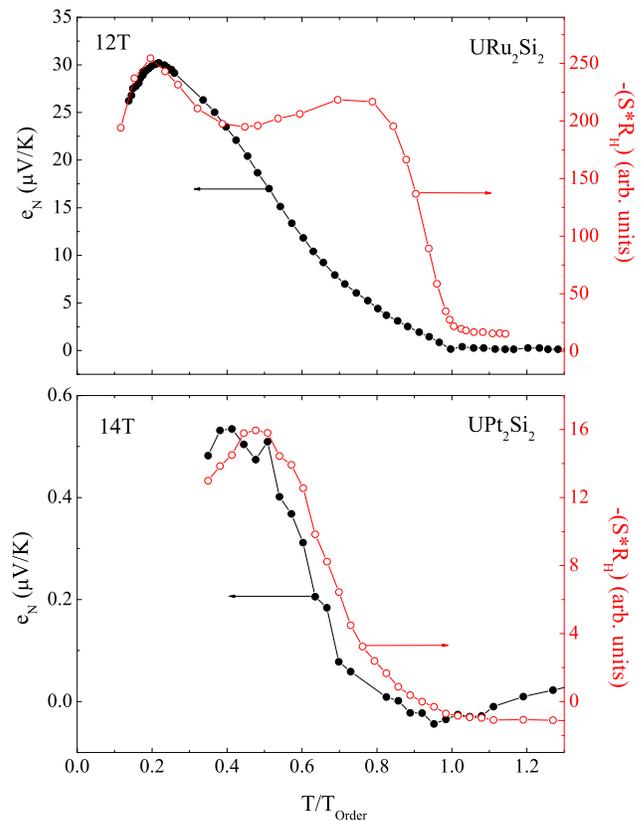}
\caption{(Color online) Upper panel: The Nernst signal of URu$_2$Si$_2$ compared to the product of its thermopower and Hall coefficient at $H = 12$\,T. Data are taken from \cite{bel04} with a $T_{Order} = T_{HO} =17.5$\,K. The lower panel depicts the scaling behavior for UPt$_2$Si$_2$ as expected from calculations for a generic antiferromagnet due to the opening of a superzone gap \cite{dik75}. The maximum positions of roughly 0.5$\cdot T_{Order}$ well agree with the theoretical expectations. Here, $T_{Order} = T_N =$\,31.5\,K.}\label{fig:eN_durchT_vs_hall_angle}
\end{figure}
  It becomes clear that the magnitude of the thermopower that might add to the transverse voltages due to a transverse temperature gradient is much smaller than the measured $e_N$. In 14\,T, $\kappa_{xy}/\kappa_{xx}$ is of the order of 1\% so that its product with the thermopower $S(T,14$T) results in a contribution of less than 10\% of the measured $e_N(T,B)$. Thus, we can safely discuss the Nernst signal as being essentially isothermal.

Using a free-electron relaxation time model, Abelskii \it et al.\rm~\cite{abelskii72,dik75} have calculated the Nernst and Seebeck effects for generic ferro- and antiferromagnetic materials. The antiferromagnets possess a superzone gap due to the doubling of the lattice periodicity. This charge gap when combined with spin and phonon scattering causes a large contribution to the Nernst and Seebeck signals below $T_N$ while the resistivity shows a much smaller effect. The calculations predict a maximum of the thermopower between 0.4 to 0.6$\cdot T_N$ \cite{abelskii72}. The Nernst coefficient is calculated in a similar manner, and is found to display also a maximum at about half $T_N$ \cite{dik75}. Within these calculations the Hall coefficient is also subjected to a steep increase after crossing into the antiferromagnetic phase. A further prediction of the calculation is a scaling of the Nernst effect with the product of the thermopower and the Hall coefficient \cite{dik75}. 

Let us compare these predictions with our measurements, starting with the thermopower in Fig.~\ref{fig:thp_temp}. The characteristic of $S(T)$ clearly displays the calculated features, a steep increase below $T_N$ with the maximum value appearing at about half $T_N$. The Nernst effect, shown in Fig.~\ref{fig:nernst_temp} also is in very good agreement with its calculated prognoses: a pronounced increase just below $T_N$ together with the maximum formation around  $0.5\cdot T_{N}$. In addition, the Hall coefficient meets the predicted behavior \cite{dik75} as well. As shown in Fig.~\ref{fig:eN_durchT_vs_hall_angle} (lower panel), the theory is once again nicely verified with respect to the scaling behavior, since the experimental data clearly display the proportionality $e_N\propto (S\cdot R_H)$. 

Figure~\ref{fig:eN_durchT_vs_hall_angle} (upper panel) displays such a scaling for the related system URu$_2$Si$_2$ ($T_{HO} = 17.5$\,K),
where it becomes evident that the underlying physics of both substances cannot be completely of the same origin. The scaling is vastly disturbed between 0.4$\cdot T_{HO}$ and $T_{HO}$. 
The differences between these two systems are obviously caused by the complicated behavior of the thermopower in that temperature region \cite{bel04}. Furthermore the positions of the maxima at $\approx0.2\cdot T_{HO}$ of $e_N$ and $S$ are shifted to lower values than the expected 0.5$\cdot T_{Order}$. One may speculate here that the HO phase is responsible for additional features besides those calculated for a generic antiferromagnet \cite{abelskii72,dik75}. 
Nevertheless, the Nernst signal of UPt$_2$Si$_2$ shown in Fig.~\ref{fig:nernst_temp} qualitatively resembles the Nernst signal found in URu$_2$Si$_2$ ($e_{N,max}(12$\,T$)\approx$\,30\,$\mu$V/K) where it develops just below the HO ordering temperature \cite{bel04}, although it is much lower in absolute values ($e_{N,max}(12$\,T$)\approx$\,0.5\,$\mu$V/K). 

In summary, we found the emergence of a large Nernst signal in UPt$_2$Si$_2$ below $T_N$, which is accompanied by anomalies in the thermopower and the Hall coefficient. These features very well agree to predictions of Refs.~\cite{abelskii72,dik75} for antiferromagnets under the assumption of a gapping of the Fermi surface below $T_N$. Our transport measurements suggest that the gap evolving in the charge channel has to be attributed to only a portion of the Fermi surface very much like that suggested for URu$_2$Si$_2$. These are common features in both related substances, URu$_2$Si$_2$ and UPt$_2$Si$_2$, but the model used for UPt$_2$Si$_2$ is not completely suitable for URu$_2$Si$_2$, indicating the influence of the itinerant character of the hidden-order phase. Nevertheless, the behavior of the Nernst effects in both systems may be connected to the gapped regions of the Fermi surface. 
\begin{acknowledgements}
This work was supported by the Deutsche Forschungsgemeinschaft through Sonderforschungsbereich 608. 
\end{acknowledgements}

\begin{thebibliography}{10}
\parskip-0.2ex plus0.05ex minus0.05ex

\bibitem{palstra85}
T.T.M. Palstra, A.A. Menovsky, J.~vandenBerg, A.J. Dirkmaat, P.H. Kes, G.J.
  Nieuwenhuys, and J.A. Mydosh.
\newblock Phys.\ Rev.\ Lett. {\bf 55}, 2727 (1985).

\bibitem{schlabitz86}
W.~Schlabitz, J.~Baumann, B.~Pollit, U.~Rauchschwalbe, H.M. Mayer, U.~Ahlheim,
  and C.D. Bredl.
\newblock Z.\ Physik B {\bf 62}, 171 (1986).

\bibitem{maple86}
M.B. Maple, J.W. Chen, Y.~Dalichaouch, T.~Kohara, C.~Rossel, M.S. Torikachvili,
  M.W.~Mc Elfresh, and J.D. Thompson.
\newblock Phys.\ Rev.\ Lett. {\bf 56}, 185 (1986).

\bibitem{broholm91}
C.~Broholm, H.~Lin, P.T. Matthews, T.E. Mason, W.J.L. Buyers, M.F. Collins,
  A.A. Menovsky, J.A. Mydosh, and J.K. Kjems.
\newblock Phys.\ Rev.\ B {\bf 43}, 12809 (1991).

\bibitem{broholm87}
C.~Broholm, J.K. Kjems, W.J.L. Buyers, P.~Matthews, T.T.M. Palstra, A.A.
  Menovsky, and J.A. Mydosh.
\newblock Phys.\ Rev.\ Lett. {\bf 58}, 1467 (1987).

\bibitem{amitsuka07}
H.~Amitsuka, K.~Matsuda, I.~Kawasaki, K.~Tenya, and M.~Yokoyama.
\newblock J.\ Magn.\ Magn.\ Mat. {\bf 310}, 214 (2007).

\bibitem{bel04}
R.~Bel, H.~Jin, K.~Behnia, J.~Flouquet, and P.~Lejay.
\newblock Phys.\ Rev.\ B {\bf 70}, 220501(R) (2004).

\bibitem{suellow08}
S.~Suellow, A.~Otop, A.~Loose, J.~Klenke, O.~Prokhnenko, R.~Feyerherm, R.W.A.
  Hendrikx, J.A. Mydosh, and H.~Amitsuka.
\newblock J. Phys. Soc. Jpn. {\bf 77}, 024708 (2008).

\bibitem{xu00}
Z.A. Xu, N.P. Ong, Y.~Wang, T.~Kakeshita, and S.~Uchida.
\newblock Nature {\bf 406}, 486 (2000).

\bibitem{wang06}
Y.~Wang, L.~Li, and N.P. Ong.
\newblock Phys.\ Rev.\ B {\bf 73}, 024510 (2006).

\bibitem{johannsen07}
N.~Johannsen, T.~Wolf, A.V. Sologubenko, T.~Lorenz, A.~Freimuth, and J.A.
  Mydosh.
\newblock Phys.\ Rev.\ B {\bf 76}, 020512(R) (2007).

\bibitem{bel04a}
R.~Bel, K.~Behnia, Y.~Nakajima, K.~Izawa, Y.~Matsuda, H.~Shishido, R.~Settai,
  and Y.~Onuki.
\newblock Phys.\ Rev.\ Lett. {\bf 92}, 217002 (2004).

\bibitem{onose07}
Y.~Onose, L.~Li, C.~Petrovic, and N.P. Ong.
\newblock EPL {\bf 79}, 17006 (2007).

\bibitem{behnia07a}
K.~Behnia, M.A. Measson, and Y.~Kopelevich.
\newblock Phys.\ Rev.\ Lett. {\bf 98}, 076603 (2007).

\bibitem{miyasato07}
T.~Miyasato, N.~Abe, T.~Fujii, A.~Asamitsu, S.~Onoda, Y.~Onose, N.~Nagaosa, and
  Y.~Tokura.
\newblock Phys.\ Rev.\ Lett. {\bf 99}, 086602 (2007).

\bibitem{lee04}
P.A. Lee.
\newblock Physica C {\bf 408-10}, 5 (2004).

\bibitem{taguchi01}
Y.~Taguchi, Y.~Oohara, H.~Yoshizawa, N.~Nagaosa, and Y.~Tokura.
\newblock Science {\bf 291}, 2573 (2001).

\bibitem{nam07}
M.S. Nam, A.~Ardavan, S.J. Blundell, and J.A. Schlueter.
\newblock Nature {\bf 449}, 584 (2007).

\bibitem{wu03}
W.~Wu, I.J. Lee, and P.M. Chaikin.
\newblock Phys.\ Rev.\ Lett. {\bf 91}, 056601 (2003).

\bibitem{berger70}
L.~Berger.
\newblock Phys.\ Rev.\ B {\bf 2}, 4559 (1970).

\bibitem{smit55}
J.~Smit.
\newblock Physica {\bf 21}, 877 (1955).

\bibitem{oganesyan04}
V.~Oganesyan and I.~Ussishkin.
\newblock Phys.\ Rev.\ B {\bf 70}, 054503 (2004).

\bibitem{dik75}
E.G. Dik and S.S. Abelskii.
\newblock Soviet Physics - Solid State {\bf 17}, 454 (1975).

\bibitem{abelskii72}
S.S. Abelskii and Y.P. Irkhin.
\newblock Soviet Physics - Solid State {\bf 13}, 2035 (1972).

\bibitem{jeffries07}
J.R. Jeffries, N.P. Butch, B.T. Yukich, and M.B. Maple.
\newblock Phys.\ Rev.\ Lett. {\bf 99}, 217207 (2007).

\bibitem{jo07}
Y.J. Jo, L.~Balicas, C.~Capan, K.~Behnia, P.~Lejay, J.~Flouquet, J.A. Mydosh,
  and P.~Schlottmann.
\newblock Phys.\ Rev.\ Lett. {\bf 98}, 166404 (2007).

\bibitem{steeman90}
R.A. Steeman, E.~Frikkee, S.A.M. Mentink, A.A. Menovsky, G.J. Nieuwenhuys, and
  J.A. Mydosh.
\newblock J.\ Phys. -- Condens.\ Matter {\bf 2}, 4059 (1990).

\bibitem{andersen80}
N.H. Andersen.
\newblock {\em Crystalline Electric Field and Structural Effects in f-electron
  Systems}.
\newblock Plenum, New York (1980).

\bibitem{berggold07}
K.~Berggold, J.~Baier, D.~Meier, J.A. Mydosh, T.~Lorenz, J.~Hemberger,
  A.~Balbashov, N.~Aliouane, and D.N. Argyriou.
\newblock Phys.\ Rev.\ B {\bf 76}, 094418 (2007).

\end{thebibliography}

*Electronic address: johannse@ph2.uni-koeln.de

\end{document}